\documentclass[prl,twocolumn,superscriptaddress]{revtex4-1}
\usepackage{amsfonts,amsmath,graphicx,amssymb}
\pdfoutput=1

\usepackage[x11names,svgnames]{xcolor}

\newcommand\kv{\mathbf{k}}

\newcommand\etal{\textit{et al.}}
\newcommand\om{\omega}

\begin{document}

\title{Absence of spin liquid in non-frustrated correlated systems}

\author{S. R. Hassan}
\affiliation{The Institute of Mathematical Sciences, C.I.T. Campus, Chennai 600 113, India}
\author{David S\'en\'echal}
\affiliation{D\'epartment de Physique and RQMP, Universit\'e de Sherbrooke, Sherbrooke, Qu\'ebec, Canada J1K 2R1}

\date{\today}

\begin{abstract}
The question of the existence of a spin liquid state in the half-filled Hubbard model on the honeycomb (aka graphene) lattice is revisited. The Variational Cluster Approximation (VCA), the Cluster Dynamical Mean Field Theory (CDMFT) and the Cluster Dynamical Impurity Approximation (CDIA) are applied to various cluster systems.
Assuming that the spin liquid phase coincides with the Mott insulating phase in this non-frustrated system, we find that the Mott transition is pre-empted by a magnetic transition occuring at a lower value of the interaction $U$, and therefore the spin liquid phase does not occur. This conclusion is obtained using clusters with two bath orbitals connected to each boundary cluster site.
We argue that using a single bath orbital per boundary site is insufficient and leads to the erroneous conclusion that the system is gapped for all nonzero values of $U$. 
\end{abstract}


\maketitle


\section{Introduction}

There is currently a keen interest in materials and models that could display a {\em spin liquid} state.
Such a state is characterized by the presence of local magnetic moments that do not order at any temperature; 
spin-spin correlations then decay exponentially as a function of distance (or algebraically, in so-called algebraic spin liquids).
In theoretical language, it can be described as an insulator that is not adiabatically connected to a band insulator, but is rather a pure Mott insulator, without spontaneously broken spatial or spin symmetries.

A spin liquid could arise from the presence of strong geometric frustration, for instance in materials with a structure resembling that of the Kagome lattice, such as Herbertsmithite, or other kind of frustrated geometries.
It has been conjectured that spin liquids could also appear in the intermediate-coupling regime of strongly-correlated systems, somewhere between a metallic (or semi-metallic) phase and a magnetic phase.
It is the latter possibility that we will entertain in this work.

More specifically, we will address the controversy about the existence of a spin liquid in the phase diagram of the half-filled Hubbard model on the honeycomb lattice. The corresponding Hamiltonian reads
\begin{equation}\label{eq:Hubbard}
H= -t\sum_{\langle ij\rangle} \left(c^\dagger_{i\sigma} c_{j\sigma} + \mathrm{H.c.}\right) +U\sum_i~n_{i\uparrow}n_{i\downarrow}
\end{equation}
where $c_{i\sigma}$ annihilates a fermion of spin projection $\sigma=\uparrow,\downarrow$ at site $i$, $n_\sigma\equiv c^\dagger_\sigma c_\sigma$ is the number of fermions of spin $\sigma$ at site $i$, and $\langle ij\rangle$ denotes the nearest-neighbor pairs on the lattice. 
This model attempts minimally at describing electron-electron interactions in a material like graphene, although a realistic description of graphene should involve long-range Coulomb interactions, phonons, and so on.
On the other hand, systems of ultracold atoms could be arranged to be fairly accurately described by the Hamilonian \eqref{eq:Hubbard}, with adjustable interaction strength $U$.

In a recent work on the matter, Meng \etal \cite{Meng:2010kx}, using Quantum Monte Carlo simulations, have argued that a spin liquid phase exists in Model \eqref{eq:Hubbard} in the range $3.5< U < 4.3$. Below $U\approx 3.5$, the model is in a semi-metallic state, and beyond $U\approx 4.3$ is is in a antiferromagnetic state (the two sublattices carrying opposite magnetizations).
But more recently, this result has been challenged by Sorella \etal\cite{Sorella:2012fk}, also using Quantum Monte Carlo simulations, albeit with larger systems.
Although this controversy seems a rather technical one, rooted in Monte Carlo methods, it also shows that the model in question, if not in a spin liquid state, is very close to one.

The presence of a spin liquid phase has been supported by other works\cite{Yu:2011fk,Wu:2012fk,Seki:2012uq} using quantum cluster methods \cite{Maier:2005}, such as the Variational Cluster Approximation (VCA) \cite{Potthoff:2003,Potthoff:2012fk}, the Cluster Dynamical Impurity Approximation (CDIA)\cite{Balzer:2009kl,Potthoff:2012fk} and Cluster Dynamical Mean Field Theory (CDMFT)\cite{Lichtenstein:2000vn,Kotliar:2001,Liebsch:2012fk,Senechal:2012kx}.
Quantum cluster methods have been used extensively in the last decade to refine our understanding of the Mott-Hubbard transition and of competing orders (magnetism, superconductivity) in strongly correlated materials. They are based on some representation of the full systems by a small, finite cluster of sites, with additional uncorrelated orbitals (the `bath') and/or adjustable source terms in the Hamiltonian.
These additional elements are determined either by a self-consistency or by a variational principle.\footnote{See supplementary material at http://link.aps.org/supplemental/*** for (i) a brief review of the quantum cluser methods used in this work, (ii) an additional numerical argument for the metallic or insulating character of the systems studied (iii) sample spectral functions for the system \textsf{h4-6b}.}
More specifically, Yu \etal \cite{Yu:2011fk} have studied the question within the Kane-Mele model, which reduces to Model \eqref{eq:Hubbard} in the special case $\lambda=0$. 
They support the existence of a spin liquid phase in the range $3\lesssim U\lesssim 4$, based on the computation of the spectral function and the associated spectral gap within VCA. Wu\etal \cite{Wu:2012fk} conclude similarly with CDMFT using a QMC solver.
Seki and Ohta \cite{Seki:2012uq} use the VCA and CDIA to study the specific question of the antiferromagnetic transition and the metal-insulator transition in Model \eqref{eq:Hubbard}.
They conclude that the single-particle gap opens up at an infinitesimal value of $U$. This also supports the existence of a 
spin liquid state.

\begin{figure}[tbh]
\includegraphics[scale=0.8]{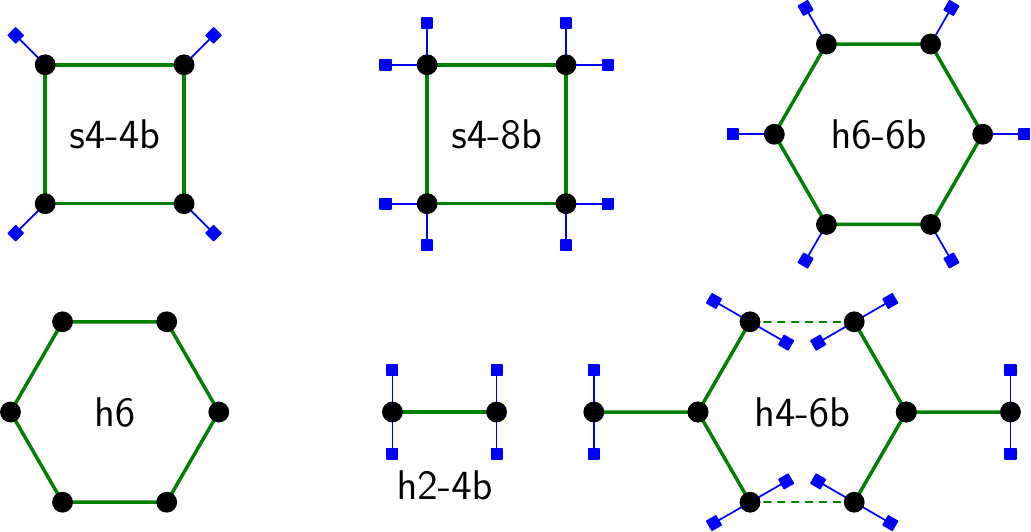}
\caption{(Color online) Clusters used in this paper. The first two pertain to the square lattice Hubbard model, the
other to the honeycomb lattice.
Blue squares represent bath sites, black circles cluster sites per se.
Dashed lines represent inter-cluster links when more than one cluster make up the repeated unit of the super-lattice.}
\label{fig:clusters}
\end{figure}

In this work we will argue, on the contrary, that Model~\eqref{eq:Hubbard} does not lead to a spin liquid phase in the intermediate coupling regime.
Instead, the transition towards a spin liquid is pre-empted by a magnetic transition towards an antiferromagnetic state, like
on the square lattice.
We will also use quantum cluster methods with an exact diagonalization solver (mostly CDMFT and CDIA), except that larger bath systems will be used.
Indeed, we assert that probing the Mott transition with a bath system of insufficient size may lead to the wrong conclusion.

The square and honeycomb lattice are both bipartite, and the half-filled Hubbard model defined on both lattices benefits from
particle-hole symmetry, which sets the value of the chemical potential $\mu$ to $U/2$, and imposes constraints on the bath parameters used in CDMFT and CDIA.
We shall therefore start with a discussion of the square lattice model, in which the same methodological issues arise, in order to put the honeycomb lattice results in perspective.

\begin{figure}[tbh]
\includegraphics[scale=0.8]{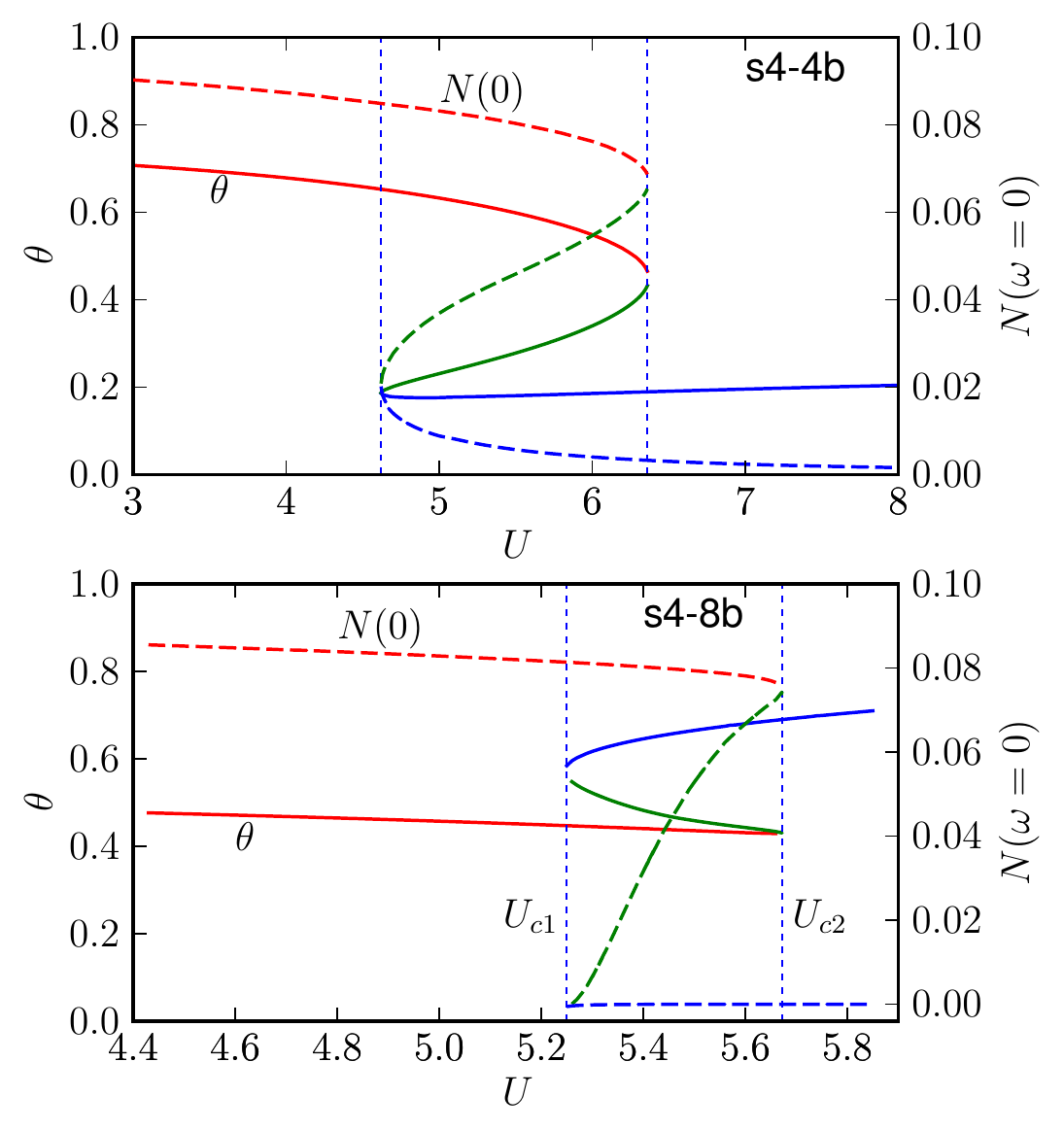}
\caption{(Color online) Full line: Hybridization parameter $\theta$ as a function of $U$, obtained in CDIA for System \textsf{s4-4b} (top) and \textsf{s4-8b} (bottom). The three solutions (metallic, unstable, insulating) are shown. The vertical dashed lines indicate $U_{c1}$ and $U_{c2}$. Dashed curve: density of states $N(0)$. Note the non-vanishing DoS in the ``insulating'' solution of \textsf{s4-4b}.}
\label{fig:s4-4b}
\end{figure}


In the square-lattice Hubbard model at half-filling, it is well-known that the Mott transition is pre-empted by the onset of antiferromagnetic order.
Nevertheless, the Mott transition may be investigated by quantum cluster methods if AF order is not allowed to set in.
This is how it was observed in Ref.~\onlinecite{Balzer:2009kl}.
In that paper, the systems \textsf{s4-4b} and \textsf{s4-8b} (see Fig.~\ref{fig:clusters}) were treated with CDIA.
Particle-hole symmetry left only one independent variational parameter in \textsf{s4-4b}: the cluster-bath hybridization parameter $\theta$. In System \textsf{s4-8b}, a bath energy $\pm\varepsilon$ was also introduced, with opposite signs on the two bath orbitals linked to the same cluster site, so as to reflect particle-hole symmetry.
A clear first-order Mott transition, with metallic, insulating, and unstable solutions, was found for the two systems, as shown in Fig.~\ref{fig:s4-4b}. The values of $U_{c1}$ and $U_{c2}$ are shown on these figures by vertical dashed lines.

To ascertain the presence of a spectral gap, we proceed as follows:
The density of states $N(\omega)$ can be computed by numerically integrating the spectral function $A(\kv,\omega)$ over wavevectors. We compute $N(\omega+i\eta)$ at $\omega=0$ for  a few values of the Lorenzian broadening $\eta$ and extrapolate $\eta\to0$ using a polynomial fit. The result of this extrapolation should vanish in the insulating solution, but not in the metallic solution.
This extrapolated density of states is shown (dashed curves) in Fig.~\ref{fig:s4-4b}.
The remarkable feature is that is does not vanish in the insulating solution associated with the system \textsf{s4-4b}, but does, as it should, in the large bath system \textsf{s4-8b}.
Thus, even though System \textsf{s4-4b} displays the a first-order transition that has all the appearances of a Mott transition, its spectral function in the strong-coupling phase has no gap, and thus this system does not adequately describe a Mott insulator.
In the context of CDMFT or CDIA, this is related to the presence of a single bath orbital per cluster site, and to particle-hole symmetry. The latter forces the bath energy $\varepsilon$ to vanish.
A correct description of the insulating state requires rather a minimum of two bath orbitals per cluster site, with equal and opposite bath energies $\pm\varepsilon$.

The presence or not of a gap may also be ascertained by computing the particle density $n(\mu)$ around the particle-hole symmetric point $\mu=U/2$ and to look
for a plateau in $\mu$, which would be the signature of a gap (the constraints on bath parameters stemming from particle-hole symmetry must then be released). This method does not require a Lorenzian broadening, as it involves integrals carried along the imaginary frequency axis. We have used both this method and the extrapolation method described above and find the same conclusions (see supplementary material).

The lesson to carry from the square lattice is that a sound description of the Mott transition (hence of a putative spin liquid state) is to be found in a cluster system with minimally two bath orbitals per cluster site.

	
\begin{figure}[tbh]
\includegraphics[scale=0.8]{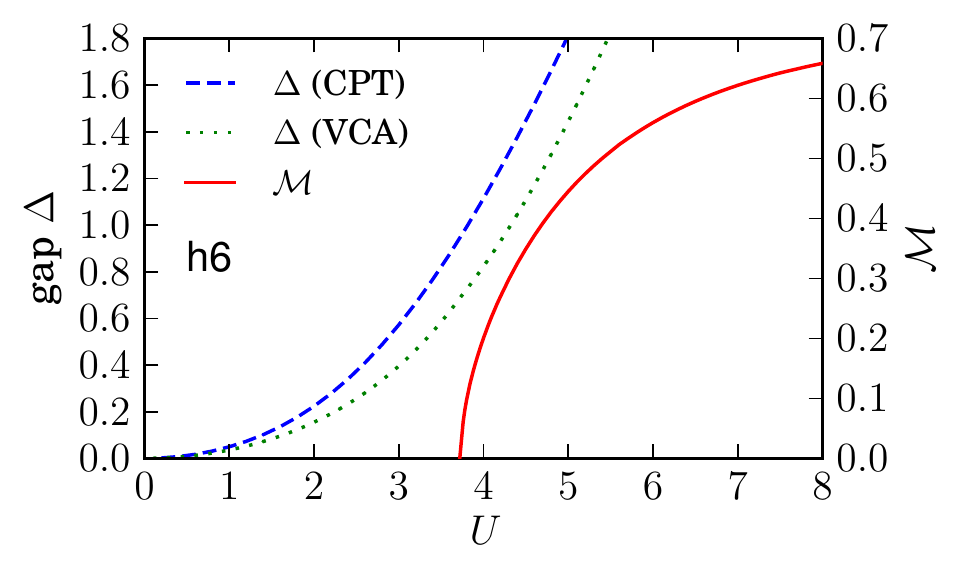}
\caption{(Color online) Result of VCA calculations on the 6-site cluster \textsf{h6}. The order parameter (red continuous curve) vanishes below $U=3.75$. Blue dashed curve: spectral gap $\Delta$ computed without variational parameters (CPT).
Green dotted curve: $\Delta$ computed from the VCA solution ($t'$ used as variational parameter).}
\label{fig:h6}
\end{figure}

Let us now turn to the honeycomb lattice.
Note that the antiferromagnetic order, on that lattice, is at zero wavevector.
Thus the nesting condition is always satisfied and we expect the AF state to be fully gapped.
This is observed in all systems studied (e.g., by inspecting the spectral function).

The first system studied is a 6-site cluster (\textsf{h6} on Fig.~\ref{fig:clusters}). It has been treated with VCA, using the nearest-neighbor hopping parameter $t'$ appearing in the cluster Hamiltonian $H'$ and a staggered magnetization $M$ as variational parameters (see supplementary material for a brief summary of the method).
As shown on Fig.~\ref{fig:h6}, the system develops a nonzero staggered magnetization $\mathcal{M}$ for $U> U_N=3.75t$.
This value of $U_N$ is remarkably close to the one found in Ref.~\onlinecite{Sorella:2012fk} ($3.75<U_N<3.8$).
This is most likely a happy coincidence, as $U_N$ will depend on cluster size.
However, the spectral gap obtained from the poles of the Green function is nonzero for all values of $U$, even those below $U_N$.
This agrees with Ref.~\onlinecite{Seki:2012uq}.
This seems a signature of a spin liquid state, but, as we will see below, systems that are better equiped to describe the Mott transition will lead us to the opposite conclusion.
Note that the gap computed from the VCA solution lies below the one obtained from the Green function without variational parameters, hinting that a better variational solution prefers a smaller gap. 

\begin{figure}[tbh]
\includegraphics[scale=0.8]{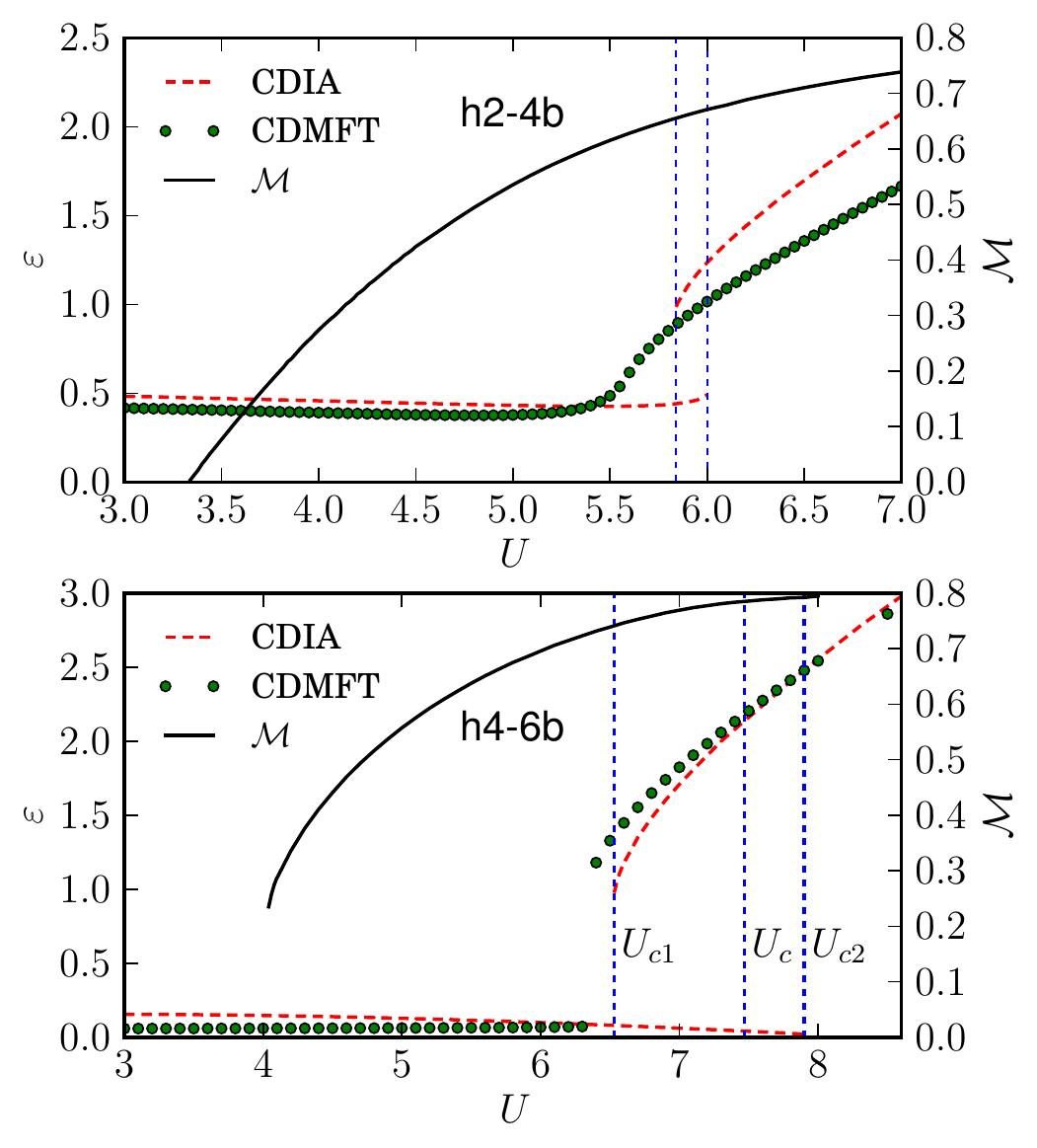}
\caption{(Color online) Top: Solution of the \textsf{h2-4b} system. Bath energy $\varepsilon$ obtained in CDIA ($U_{c1}$ and $U_{c2}$ are indicated by dotted vertical lines) and CDMFT. The staggered magnetization $\mathcal{M}$ obtained by CDIA is also shown.
Bottom: same, for System \textsf{h4-6b}. The value $U_c$ where the metallic and insulating solutions have the same energy, is also indicated.}
\label{fig:h2-4b}
\end{figure}

System \textsf{h6-6b}, with one bath orbital per cluster site, was also studied, and our calculations agree with those of Ref.~\onlinecite{Seki:2012uq}: the system has a spectral gap for all values of $U$ and displays no Mott-insulator transition (see supplementary material).
However, we assert that probing the Mott transition in this system is unreliable, just as it is in System \textsf{s4-4b} for the square lattice. It is safer to use systems with two bath orbitals per cluster site.
The simplest such system for the honeycomb lattice is \textsf{h2-4b} (Fig.~\ref{fig:clusters}).
We studied this system both with CDMFT and CDIA. At half-filling, particle-hole symmetry demands that the bath energies
of the two bath orbitals connected to the same cluster site be opposite in value ($\pm\varepsilon$).
In the non-magnetic state, the two sites of the cluster (and the correponding bath sites) are related by left-right symmetry, and
therefore only two bath parameters remain: one bath energy $\varepsilon$ and one hybridization parameter $\theta$.
The CDMFT paramagnetic solution for $\varepsilon$ is shown on the upper panel of Fig.~\ref{fig:h2-4b}.
An upturn in the value of $\varepsilon$ at around $U=5.5$ signals the Mott transition.
But no hysteresis is seen when the interaction $U$ is swept upwards and downwards, which means that CDMFT in this case does not detect the first-order character of the Mott transition.

\begin{figure}[tbh]
\includegraphics[width=0.9\hsize]{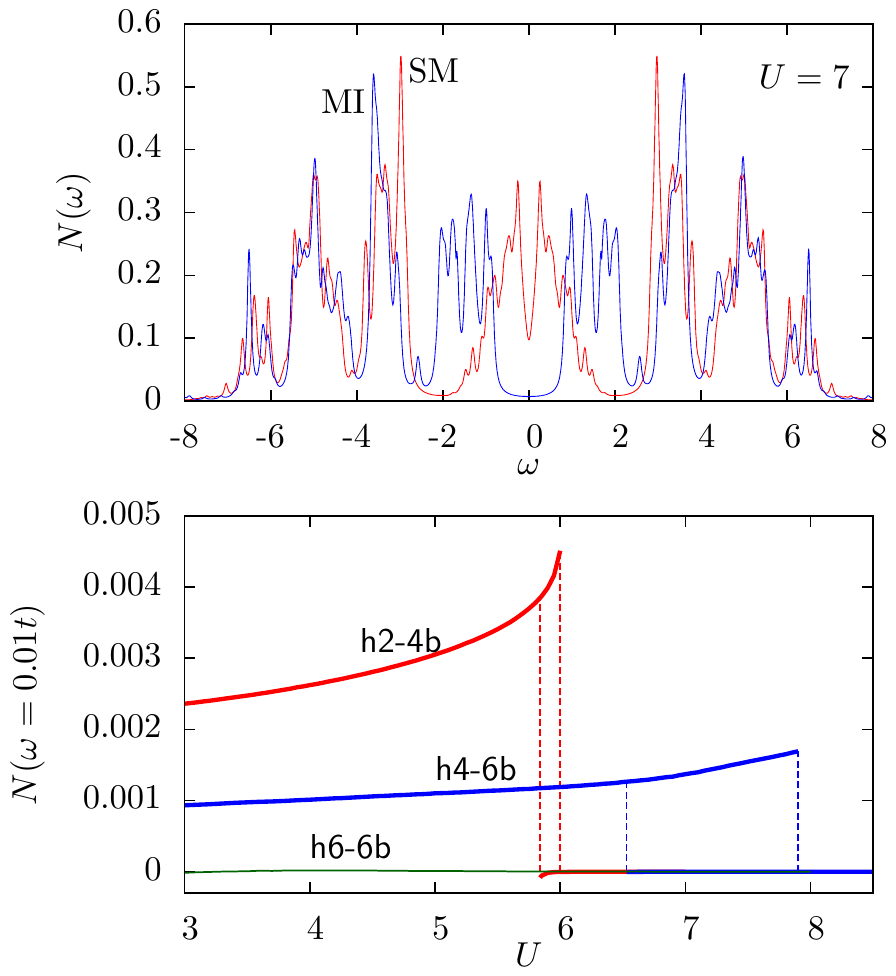}
\caption{(Color online) Top panel: density of states in the semi-metallic (SM, red) and insulating (MI, blue) solutions obtained from System \textsf{h4-6b} at $U=7$, with a Lorenzian broadening $\eta=0.05t$. Bottom panel: density of states $N(\om=0.01t+i\eta)$, extrapolated to $\eta\to0$ for different systems, as indicated. Only the semi-metallic solutions of \textsf{h2-4b} and \textsf{h4-6b} have a non negligible value.}
\label{fig:dos0}
\end{figure}

Things are different when CDIA is applied to the same system. As shown again on the upper panel of Fig.~\ref{fig:h2-4b}, two CDIA solutions are found: a semi-metallic solution when $U$ is increased and an insulating solution when $U$ is decreased. Each of these stops, respectively at $U_{c2}$ and $U_{c1}$, and coexist in the range $[U_{c1},U_{c2}]$.
The first-order character of the solution is therefore clearly seen in CDIA.
Using the same extrapolating method used in the square lattice case, it is easily verified that the spectral gap vanishes throughout the semi-metallic solution, whereas it is nonzero in the insulating solution (see Fig.~\ref{fig:dos0}).
Note that in that case, $N(\omega+i\eta)$ is computed at $\om=0.01t$ instead of $\omega=0$, since $N(0)$ is expected to vanish in the semi-metallic phase.

Also shown on Fig.~\ref{fig:h2-4b} is the antiferromagnetic order parameter obtained if the Weiss field $M$ is added to the list of variational parameters; this constitutes in fact a mixture of CDIA and VCA, since the parameters at play within Potthoff's variational approach are both bath-related ($\varepsilon$ and $\theta$) and cluster-related ($M$).
We find that, in this system, the critical interaction strength for the onset of magnetic order is $U\approx 3.4$, an even smaller value than in the 6-site cluster VCA computation. Thus in this system the (continuous) magnetic transition pre-empts the Mott transition and no spin liquid occurs.

The larger cluster system \textsf{h4-6b} (see Fig.~\ref{fig:clusters}) with two bath orbitals per site was also studied.
In this case, two 4-site clusters are necessary to form a repeated unit.
Each cluster is hybridized to 6 bath orbitals, two on each edge site (the central site is not coupled to the bath).
Again, particle-hole symmetry and rotational symmetry make for only two independent bath parameters, $\varepsilon$ and $\theta$, like for the smaller system \textsf{h2-4b}.
The solution is shown on the lower panel of Fig.~\ref{fig:h2-4b}.
When CDMFT is applied, the Mott transition appears clearly at $U\approx 6.35$, but no hysteresis is observed.
Again, CDIA finds a semi-metallic and an insulating (spin liquid) solution, which overlap between $U_{c1}$ and $U_{c2}$.
Their energy $\Omega$ are equal at an intermediate value $U_c$ (indicated on the figure).
Like in the case of the system \textsf{h2-4b}, the spectral gap vanishes in the semi-metallic phase (Fig.~\ref{fig:dos0}).
If the cluster Weiss field $M$ is added to the list of variational parameters, the CDIA predicts an antiferromagnetic transition at $U_N=4.0$, which again means that the Mott transition is pre-empted. Thus, this larger system also rules out a spin liquid phase.

In conclusion: We have applied various quantum cluster methods to the Hubbard model on a honeycomb lattice, in order to investigate
the possible emergence of a spin liquid state.
We make the hypothesis that the spin liquid state that might emerge in a strongly correlated system without magnetic frustration, such as the one studied here, coincides with the Mott insulating state. 
The Mott transition itself cannot be adequately accounted for by CPT or VCA: the cluster's environment must be described by a bath of uncorrelated orbitals, i.e., by a dynamical mean field, and this bath must be large enough (two bath orbitals connected to a single cluster site).
This leaves CDMFT or CDIA as adequate cluster methods to study the Mott transition. Two nonmagnetic solutions (a semi-metal and an insulator, aka spin liquid) are found, separated by a first-order transition. The CDIA is the better approach, since it reveals clearly the three critical values $U_{c1}$, $U_{c2}$ and $U_{c}$. The spectral functions computed from these solutions show the persistence of the Dirac cones up to the Mott transition, hence the gapless character of the semi-metallic solution. The magnetic solution can also be obtained in CDIA, and always appears at a much lower value of $U$ than the Mott transition. This leads us to assert that a spin liquid (aka Mott insulator) does not exist in this system: it is pre-empted by magnetic order. The critical value of $U$ at which the magnetic solution appears is comparable to what is found in large-scale Monte Carlo simulations\cite{Meng:2010kx,Sorella:2012fk}. Since the Mott transition is a rather local phenomenon, we argue that increasing the cluster size, which we cannot do with our exact diagonalization solvers, would not affect the value of $U_c$ to the point of changing our conclusion.\cite{[{Indeed, a variational study of the Mott transition in the one-dimensional Hubbard model with NN and NNN hopping has shown that $U_c$ is rather independent of system size, even though there is a size-effect in the fluctuations surrounding the transition. Note that in CDMFT and CDIA the fluctuations are temporal more than spatial, whereas temporal fluctuations are not taken into account in a variational wavefunction approach. See }][]Capello:2005fk}
In fact, increasing the cluster size would likely shift $U_c$ to a slightly higher value.
Therefore we believe that our conclusion carries over to large clusters.


We thank G. Baskaran, J.-P. Faye, R. Shankar, P.V. Sriluckshmy and A.-M. Tremblay for useful discussions.
Computational resources were provided by Compute Canada and Calcul Qu\'ebec.


%

\end{document}